\documentclass[10pt,amssymb,aps,pre,preprint]{revtex4-2}
\usepackage{graphicx}
\usepackage{hyperref}
\usepackage{url}
\usepackage{setspace}
\usepackage{xcolor}
\usepackage{epstopdf}
\usepackage{amsmath}

\begin{document}
\title{Force chain dynamics in a quasi-static granular pile}
\author{Benjamin Allen, Nicholas Hayman}
\affiliation{$1$ Oklahoma Geological Survey, University of Oklahoma, Norman OK 73071}

\date{\today}

\begin{abstract}
In nature, granular materials fail in abrupt avalanches, earthquakes, and other hazardous events, and also creep over time. Proposed failure mechanisms for these systems are broadly framed as friction-limited.  However, mechanical descriptions of friction in granular system vary, including those that consider the non-linear, heterogeneous dynamics of grain-contact forces. In order to study granular controls on failure and creep, we imaged contact forces between quasi-2D photoelastic discs at 15-minute intervals in experiments over weeks- to one-month-long observation periods. In the experiments, the particles are distributed in a slope below the angle of repose with a concomitant establishment of the force-chain network approximately along the principal static stress directions.
The discrete force-chain shifts are initially described by an age-weakening Weibull distribution in frequency over time, with the most common clusters of events $<$1000 minutes apart, but there are long quiescent periods that are not well described by the distribution. Different surface slopes show the same contact aging rate, but the steeper the slope the more likely there is to be a failure event. The post-settling discrete events accompany a longer term strengthening of the localized stresses. We observe that some events may be related to 2$^o$C temperature change, but associated ground motions measured by a laboratory-installed seismometer appear to have no correlation with particle displacements or force-chain changes.  The sum of all ground motion is four orders of magnitude smaller than the temperature changes, further ruling out mechanical noise as an appreciable cause of events. The results illustrate that local, grain-scale changes in force-chain networks can occur long after a granular pile reaches a mesoscale apparently stable state, sometimes without obvious external forcings or imposed state changes. Such force-chain dynamics may underlie transitions in natural granular systems between large-scale failures and creep events.
\end{abstract}

\maketitle

\section{Introduction}

Work at the intersection of geosciences and physics suggests that many hazardous failure events such as landslides, avalanches, and even earthquakes occur as grains become unjammed in complex ways \cite{jerolmack2019viewing, daniels2008}. In the study of natural slope failure and creep, a common framework employs empirical friction laws that describe a quasi-stable regime governed by the effects of fluid pressure and frictional stability \cite{iverson2005regulation, finnegan2024seasonal}. The stability of faults, and their propensity for earthquakes vs. creep, have a long history of characterization via experimentally derived rate-and-state friction laws \cite{dieterich1972, marone1998laboratory, nie2024velocity}. However, recent experiments further identified smooth,  diminishing slip below the static friction threshold \cite{sirorattanakul2025}. Other experiments report transient motions in passive, sloped granular piles  \cite{deshpande2023sensitivity}. These behaviors raise the question if there is a new framework needed for describing a broader range of granular behaviors. 

Granular materials behave in a manner that is not precisely that of a solid, liquid, or gas \cite{jaeger1996granular} and granular systems can fail abruptly (i.e. avalanches) but also can deform over time scales longer than most observations periods. Small, slow deformations are broadly called \textit{creep}. For granular systems, as with other failure phenomena, creep is often considered to arise through grain-scale dynamics \cite{roering1999,roering2001,jerolmack2019viewing,sharpe1938,terzaghi1950,parizek1957, allen2018, sirorattanakul2025}. Here, we further restrict our definition of creep to the non-reversible movement of particles below the initiation of rolling-sliding transport.

One way to describe granular systems is to consider both the average micromechanics of grain contacts, and the inhomogeneous stress distributions that arise as the system is driven towards a fully constrained particle configuration. Sloping granular piles are an excellent way to probe this contrast between average and inhomogeneous behaviors. At the mesoscale, a slope is considered stable when it exists below an angle of repose which is in turn determined by the geometry and friction of the particles. In nature, steeper slopes are more likely to fail via avalanches and mass wasting, but shallower slopes can also slump, slide and creep \cite{ribiere2005, ferdowsi2018}. However, the angle of repose actually has a range of values when the angle is either decreased or increased to attain the repose angle \cite{beakawi2018,cheng2017}.  Then, as a granular pile approaches a state of rest, the force distribution and particle contact chains will evolve as the system evolves from a fragile state to a fully jammed state \cite{cates1998,liu2010, bi2011, behringer2014,behringer2019}.

There are a large number of possible jammed states with different particle configurations within a small range of densities, and these states exist well before the system approaches the global equilibrium state. The nature of jamming is such that it creates structure out of a random configuration of particles including arches and pockets where particles can be held without being confined.  For example, an interesting byproduct of these configurations is that unconfined particles, or \textit {rattlers}, are easier to move but are not part of the main force bearing structure, and the kinetic energy of these particles decays slower than that of the bulk \cite{giacco2017}. Other granular structures form under stress, with and without rattlers, concentrating the particle stresses along a chain of particle which can support small deflection of the forces and arch-like structures \cite{cates1998}.  Granular structures are often described as having an average state, but the mechanics of granular structures is also determined by the fluctuations of a few constituents rather than the mean behavior of all the particles \cite{claudin97} and not all observed behavior can be described by average constitutive relationships. This makes employing granular-physics models for predictive earth-science problems challenging, but also is why there are many still undiscovered local behaviors that could underlie mesoscale behaviors, such as landslides and slope creep. 

Here, we present the results of experiments that allow the visualization of in situ stresses in a tapered granular pile, with minimal perturbation. Local stress changes and particle motions all happen below the angle of repose, and in some instances long after the pile settles into an apparently stable configuration. In order to investigate these dynamics, we take advantage of the direct imaging of particle forces within birefringent materials under cross-polarized light \cite{daniels2008,daniels2017,zadeh2019}. This method has been used to explore cooperativity between particles\cite{thomas2019force}, fault slip \cite{daniels2008, hayman10}, avalanches \cite{bares17}, impact and penetration experiments \cite{clark2012}. The resulting polarized light in the photoelastic discs allow for the direct imaging of stress localization in isotropic grain packings by illuminating \textit{force chains}, the linking of localized stress to form long-range directionally interacting particles. 

Force-chain changes in a granular system highlight the jammed stress state evolution. Unjamming a granular system by imposing shear \cite{hayman10,bi2011,pan2023,dahmen2011} or by tapping or shaking \cite{knight95,ribiere05,johnson2008}, allows particles to redistribute, followed by settling into a more ordered state. 
In some granular systems heating causes an increase in avalanche events and granular displacement, wherein mechanical noise consolidates the pile and strain localizes to the top layers of grains\citep{deshpanade2021}. Both of these effects are considered to cause catastrophic failure in grain silos\cite{janssen1895,marconi2000}. When examining the erosion of a granular bed under the effect of fluid shear, settling and lateral movement occur below the erosion threshold resulting in creep at approximately all shear rates\cite{allen2018}. Our focus here is to probe the force-chain and particle displacement dynamics in as static a system as possible, a heretofore relative unexplored systems \cite{deshpande2023sensitivity}. We present our new experimental results followed by a discussion relating the apparently fragile state of the experimental system to similar systems in nature. 

\begin{figure}
    \centering
    \includegraphics[width=0.85\columnwidth]{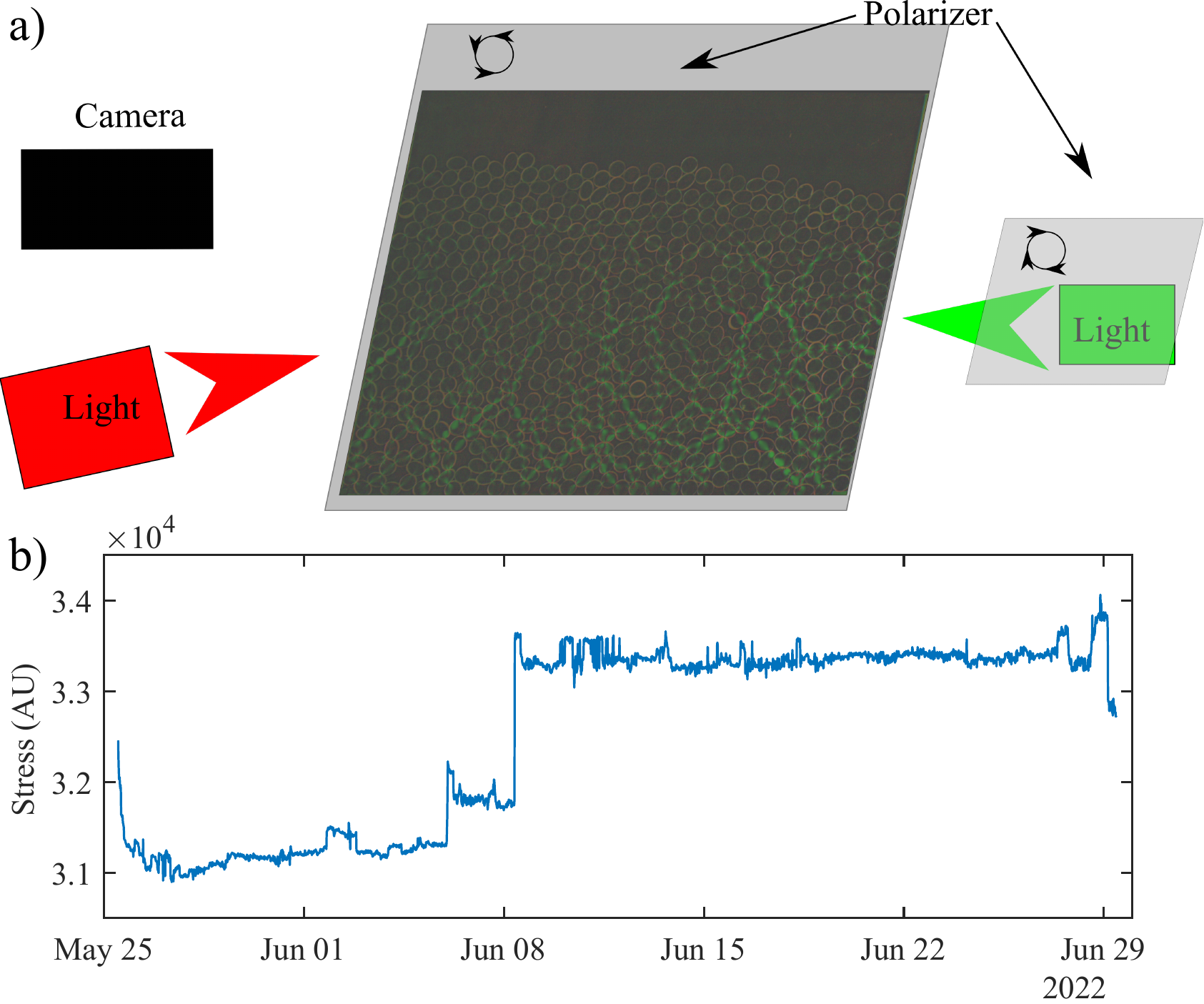}
        \caption{
        (a) The apparatus for the system.  Polarized green light shines through the system from the back, and red light reflects from the front through a second opposite polarizer into the camera.
        (b) The overall localized stress in arbitrary units based on the intensity ($\nabla I^2$)  of the force chains measured in force chains over a month with a slope of 28$^o$ shows an initial relaxation at the beginning of the experiment before the localized stress grows in discreet jumps.  Longest experiment of 35 days shown, additional experiments shown in Supplemental S5.}
    \label{fig:apparatus}
\end{figure}

\section{Experiment}

The experimental apparatus comprises two vertical 16" x 20" glass panes separated by 3/8" giving the discs 1/8" of an inch space between the glass plates.  The edges of the discs are illuminated with red light and force chains with cross-polarized green LED light as shown in Figure \ref{fig:apparatus}(a). The experiment has a population of 403 circular discs with diameter, $d_{circ}$ $=11.8$ mm and 214 elliptical disc with major axis $ax_{maj} = 14.4$ mm and minor axis $ax_{min} = 9.7$ mm with a thickness of $h=6.4$ mm for both.  The ellipses help prevent crystallization in the system, though using them comes at the expense of some quantitative control over stresses \cite{kozlowski2022}. The discs are cut from PSM-4 Vishay birefringent photoelastic material which has an elastic modulus $Y=4.14$ MPa, density $\rho=2.68 g/cm^3$, thermal expansion coefficient $\alpha \approx 0.00029$ C$^{-1}$, and Poisson ration $\nu\sim0.5$.

The experimental setup allows the direct imaging of force chains when the particle stresses are $>20$ mN, approximately 3 particle weights. As the grains are stressed, the index of refraction changes causing a shift in the polarization of light created by the birefringence. The index of refraction change can be seen with circular polarizing filters on either side of the experiment.  Circular polarization is more uniform than linear polarization which is angle dependent.  The circular polarizers are created with a quarter wave-plate and a linear polarizer, so it is important the quarter wave plates are focused towards the experiment \cite{daniels17,puckett12,majmudar06}. The resulting birefringent patterns in the photoelastic discs allow for the direct imaging of stress localization in isotropic grain packings by illuminating the force chains. 

To create the settling pile, we rotate the filled apparatus around the horizontal axis facing the camera to create a slope.  This procedure allows for gross control ($\pm 1^o$) on the slope of the experimental slope but each distribution of the particles in the slope is unique.

The experiment is placed for a few weeks in an empty basement room with an image of pixel size 250 microns x 250 microns taken every 15 minutes. For the experiment closest to the angle of repose corresponding temperature/humidity measurements and seismic monitoring are taken at complimentary temporal resolution next to the experiment to determine the effect of possible perturbations. With the seismic sensor we detect the nearby train passings, opening and closing of doors, and the building shifts and settling with the Raspberry Shake strain-rate sensor sitting on the same surface as the experiment.

\section{Results}

The experiments show an initial relaxation time where the stress redistributed with an exponential decay constant of $\tau\sim 5.5$ hours. Following the initial decay the localized stress in the force chains subsequently begins a slowly increasing trend with stress increasing and decreasing as discrete events as shown in Figure \ref{fig:apparatus}(b).  In no experiment do the particle displacement exceed half of their diameter, $<$0.5$d$. Displacements create a new stress distribution and network, but stress shifts also occur without visible particle displacement. Even after four weeks, force-chain fluctuations occur.

\begin{figure}
    \centering
    \includegraphics[width=0.95\columnwidth]{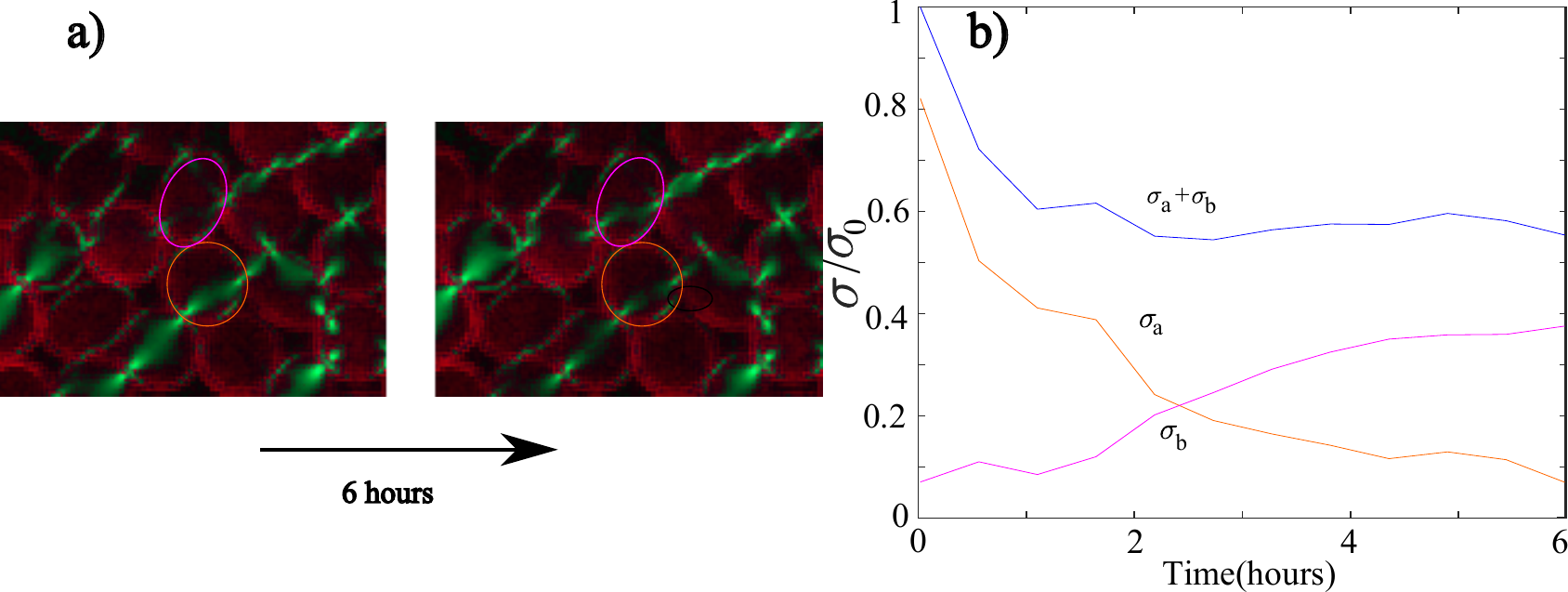}
    \caption{(a) An example of a set of force chains at the beginning of the experiment and again 6 hours later during the initial transient settling. (b) The evolution of the stresses over those 6 hours calculated with the intensity of the particles shown in (a).  Colors match between (a) and (b) and the sum of both is in blue showing that the stress distribution also spreads to other chains.}
    \label{fig:stressshift}
\end{figure}

\begin{figure}
\includegraphics[width=0.65\columnwidth]{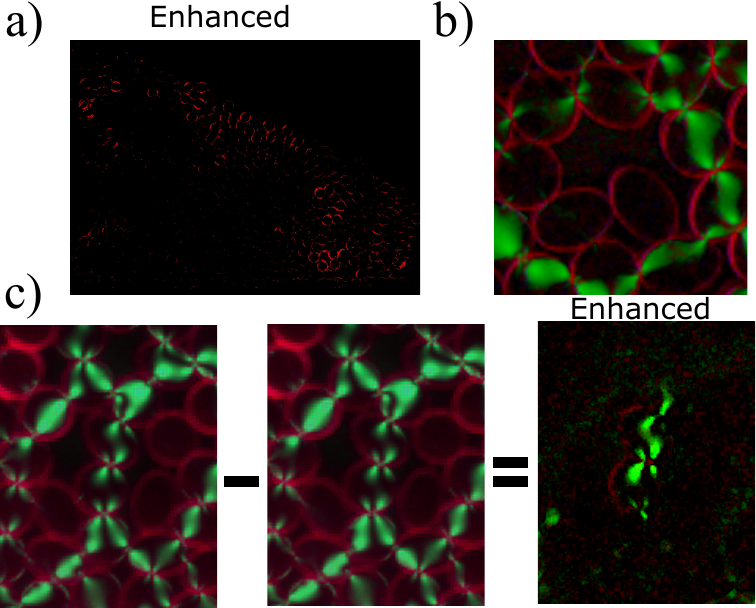}
\caption{(a) Difference Image of particles in the slope before and after 28 days.  Most of the movement occurs in the first six hours. No particle moves more than a particle diameter. (b)Rattler/Observer particle in a pocket, the particle is not fully constrained (c) Force chain flex between 323.75-324 hours shown in difference imaging. (See Supplemental Video 2)}\label{fig:dispshift}
\end{figure}

During any experiment, because of the disordered random packing of particles, forces are not evenly distributed and there are particles that are not fully confined. This state allows particles to move\cite{giacco2017} without significantly changing the overall stress network, as shown in Figure \ref{fig:dispshift}(a). At the top of the slope, with no confining pressure, the particles can break free and tumble until further confined.  Deeper within the experiment the particles move within their confining pocket.

Additionally, particles can be confined along a single direction, as shown in Figure \ref{fig:dispshift}(c).  These particles exhibit a classic example of fragile jamming \cite{cates1998} where they are confined directionally, but are weak in other directions. In these instances, the chain of particles has a stability condition determined by the friction between the particles as well as the rotational confinement.  In such a state, the chains can shift and the corresponding force chain can flex as shown in Figure \ref{fig:dispshift}(c).  This shift took place between 323.75 and 324 hours into the experiment where the slope was left to settle and shift under its own weight.

Finally, if the particles are driven out of stability there will be failure.  This failure manifests differently depending on the stress state and confinement of the particles. At the top of the experiment particles can avalanche and tumble down the slope, providing impulses and disturbances potentially taking particles lower on the slope out of stability.  This instability is described by the macroscopic angle of repose which estimates the friction of the particle against the weight of the particle and the slope below it. Deeper within the experiment, where there is confinement, the failure of force chains leads to localized failure within shear bands along the force chain direction\cite{schall2010}, this the mechanism of landslides. However, in this study we do not see any distinct shear banding failure. 

For a particle to move, it must be unconfined and a net force must be applied. Any resulting contact failure can be visualized by the force-chain shifts within the pile accompanied by particle movement. The force-chain changes then occur within the granular pile under gravity over timescales of minutes, days, and weeks. The stress fluctuations caused by the particle contact network changes are spatially heterogeneous and often shift over time without any noticeable change in particle position; this can be explained simply because friction is a non-conservative force resisting movement over a range from no shear to shear, defined by the coefficient of friction multiplied by the confining normal force. An example of stress shifting without particle displacements is shown in Figure \ref{fig:stressshift}. In this instance, over the first 6 hours of the experiment the force chain migrates upward to a different set of particles, but without shifting the particles. Over this force-chain migration, the overall stress decreases suggesting a redistribution of stresses to other particles.

\begin{figure}
    \centering
    \includegraphics[width=0.85\columnwidth]{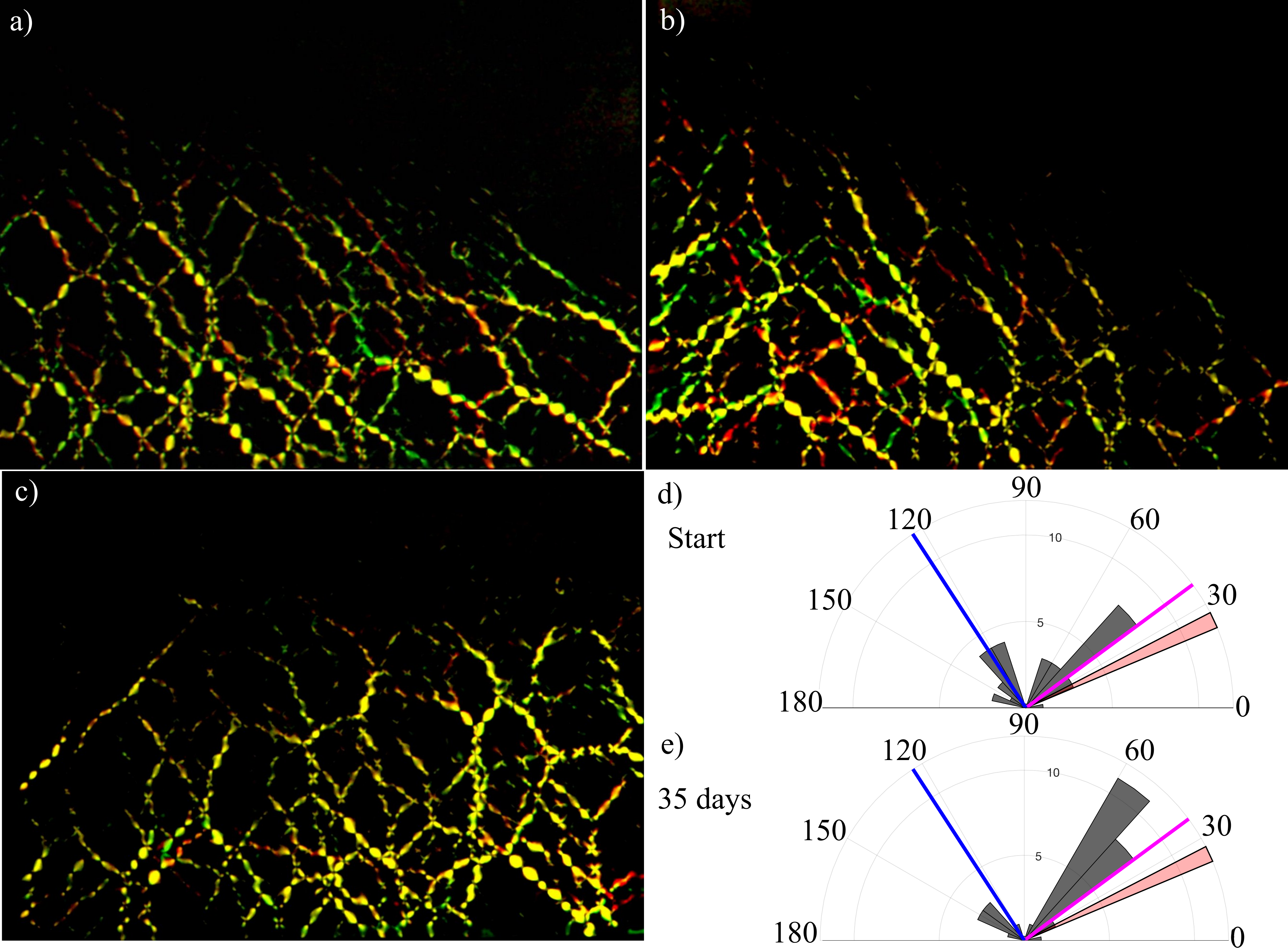}
    \caption{(a) The force chain distribution with slope 27 $^o$, the earliest time frame is in green and the final one is in red; yellow shows the force chains that do not change; (b) The force chain distribution with slope 19.5$^o$; (d) The force chain distribution with slope 7.2$^o$; (d) The initial angular distribution of the top 30 longest force chains for the longest experiment (35 days, 27 $^o$ slope). (e) The final angular distribution of the top 30 longest force chains after 35 days of the longest experiment (35 days, 27 $^o$ slope).  The blue and purple lines show the calculated principle stresses from the slope, and the shaded wedge shows the slope which does not change measurably in time.}
    \label{fig:fchainangle}
\end{figure}

Figure \ref{fig:fchainangle} shows an example of the evolution of the force chain network's preferred orientations and intensities over a month-long experiment. One can observe significant changes in the force chains, with about half of the force chains shifting or disappearing over time. The forces at the top of the pile normalize, disappearing as they fall beneath the threshold of 3-4 particle-weights and transition down-slope and deeper into the pile. The directionality of the force chains corresponds to the principle stresses of the material and also change over time, without a change in the bulk measurements (i.e. slope).

In terms of the principal stresses, as shown in Figure \ref{fig:fchainangle}(d), the angles of the 30 longest force chains in the experiment lie scattered around the calculated principle axis of the granular angle of repose $\theta\approx 28^o$ measured from the experiment.  A month later the 30 longest force chains reoriented to steeper angles, shown in Figure \ref{fig:fchainangle}(a-c) despite no apparent change in the angle of repose of the pile.  The rearrangement of the stress network suggests that over time the principle stresses in the pile are slowly changing as the pile settles\cite{allen2018} (Figure \ref{fig:fchainangle}). Such non-equilibrium behavior is also born out with the localized stress rising as time goes by, rather than falling before failing, causing a new force configuration and subsequent localized of stresses.

\begin{figure}
    \centering
    \includegraphics[width=.85\columnwidth]{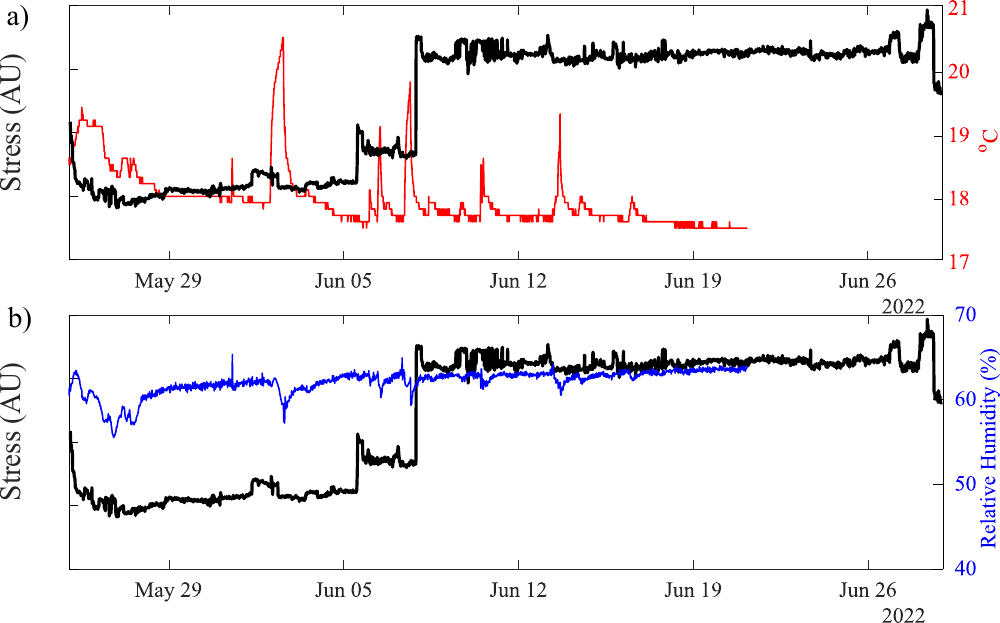}
    \caption{(a) The photoelastic stress in arbitrary units and the temperature over the course of one experiment at $\theta = 27.5^o$.  
    (b) The humidity over the course of the same experiment.}
    \label{fig:temphumid}
\end{figure}

We now turn to potential outside inputs of energy that could possibly be responsible for the disequilibrium state of the granular piles. A key environmental variable is the ambient temperature shown in Figure \ref{fig:temphumid}; note that humidity (not shown) is inversely proportional to temperature. Many $2^o$C changes result in changes in the force-chain energy and configuration as shown in Figure \ref{fig:temphumid}. Importantly, however, many temperature changes are not accompanied by force-chain changes or particle displacements, and vice versa. 

\begin{figure}
        \centering
        \includegraphics[width=0.85\columnwidth]{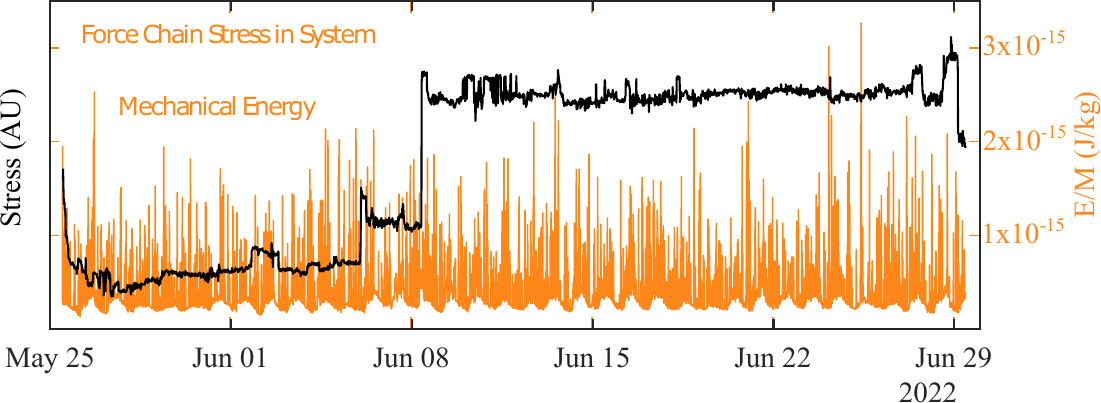}
        \caption{The overall stress in arbitrary units as measured from the intensity plotted against time.  There is an initial transient over with an exponential time constant of $\sim 5.5$ hours as the pile settles and then smaller shifts and events as the force chains slowly fluctuate and rearrange. Red lines mark 54 sharp changes in the intensity and hence the stresses, (ie force chain breaking/rearrangment).}
        \label{fig:energyinput}
    \end{figure}

In addition to the changing temperature and humidity, there is a variety of background mechanical noise potentially impacting the experiment. We measured the mechanical perturbations to the experiment surface with a Raspberry Shake 1-D geophone. Notable identifiable events are from nearby (~0.5-1.0 km away) train traffic and the settling of the building in the early morning hours. Neither of the events shows any clear correlation with the shifting of the force chains or the settling of the structure as shown in \ref{fig:energyinput}.  The induced mechanical energy is on the order of $E=mv^2\approx 10^{-15} J$ total over the 15 minute measurement periods.  Individual events and perturbations are smaller by about three orders of magnitude.

\section{Discussion}

The description of failure, creep, and related phenomena as resulting from 'glassy' transitions opens up a line of research involving a range of transitions as grains jam and unjam, and force chain networks develop and evolve  \cite{jerolmack2019viewing}. Previous experiments have recognized motions in quasi-stable experimental granular piles \cite{deshpande2023sensitivity}. However, investigation of long-term force-chain evolutions such as in our experiments have not, to our knowledge, been deeply investigated. We begin a discussion of our results by considering the energetics of the system. This has two roles: (i) to determine if the environmental noise is adding energy to the system and causing force-chain changes and particle displacements, and (ii) to gain some understanding of the energetics of failure and creep transitions from a grain-scale perspective. We finish our discussion by focusing on the apparent fragile state of our experimental system, and what might be learned about corresponding natural systems.

The energy of the system can be written as:
\begin{equation}
E_{tot} = \sum_{i}\left( E^i_g - \frac 12\sum_{j\neq i} \left(E^{ij}_{friction}+E^{ij}_{contact}\right)\right) +E_{mech}
\end{equation}
where $i$ and $j$ denote each particle. $E^i_g=m_i g z_i$ does not change in time while the particles are at rest. 
$E^{ij}_{contact}$ is the energy stored in each contact between particles $i$ and $j$.  Input energy is dissipated by the friction resisting any movement $E^{ij}_{friction}$ with the only possible storage of energy in the elastic particle contacts $E^{ij}_{contact}$.  Finally, there is $E_{mech}$ which is the determination of any ambient noise from the environment.  The balance of energies ($E_g$ and $E_{contact}^{ij}$) differs between slopes, but the respective energy is zero when the particle, and system, is at rest.  
The discrete changes in the system occur from changes in the forces with time.  If those forces become unbalanced, then work is done on the system, by changing particle positions and thereby changing the energy state.

In order to explore the potential impact of added mechanical energy from things like regional noise, we assume that any mechanical energy added to the system can be represented with $E_{KE}=Mv^2$ and integrated for each 15 minutes to match the force chain measurements.  Dissipation is ignored for the calculations of mechanical energy. The mechanical energy interjected into the system within 15 minutes varies from between 7 $\times 10^{-15}$ to 30 $\times 10^{-15}$ Joules per particle and over the course of the 35 day experiment totals 2.8$\times 10^{-10}$ Joules per particle.  These energies, which are integrated over 15 minute time scales, represent a vertical movement of the particles by 1-4 picometers against gravity, assuming the energy is changed directly into movement. Therefore, from a strictly energetic standpoint, mechanical noise is not a particularly strong driver of any particle motions or force-chain behaviors in our experiments. 
 
Temperature changes could be a more likely energy input. The corresponding energy from the temperature is approximately a maximum of a Joule for the measured 2$^o$ change of temperature, assuming a specific heat capacity of $0.3-2$ Joules g$^{-1}$ C$^o$. This also assumes and that we can use the constant pressure heat capacity where $E_T=c_pM\Delta T$. Figure \ref{fig:energyinput}(a) shows the temperature signal which might correlate with some of the events, though not all of them. In contrast, the mechanical energy only shows a couple small stress changes that can be correlated with events shown in Figure \ref{fig:energyinput}(b),

Not all of the force chain shifts correspond to temperature changes but temperature changes can cause a small change in the size of the particles where $d = d_0+\alpha d T$. The radial expansion of the particles during these 2$^o$ C temperature spikes adds a stress of $\sim 20$ $\mu$J to any adjacent particle that can not expand.
This stress change is 100x less than adding another particle to the top of the slope.  Thus, the energy input from temperature changes is very small, but because of the ways it is accommodated, there is reason to consider apparent correlations between temperature changes and force-chain changes or particle displacements.

\begin{figure}
    \centering
    \includegraphics[width=0.95\columnwidth]{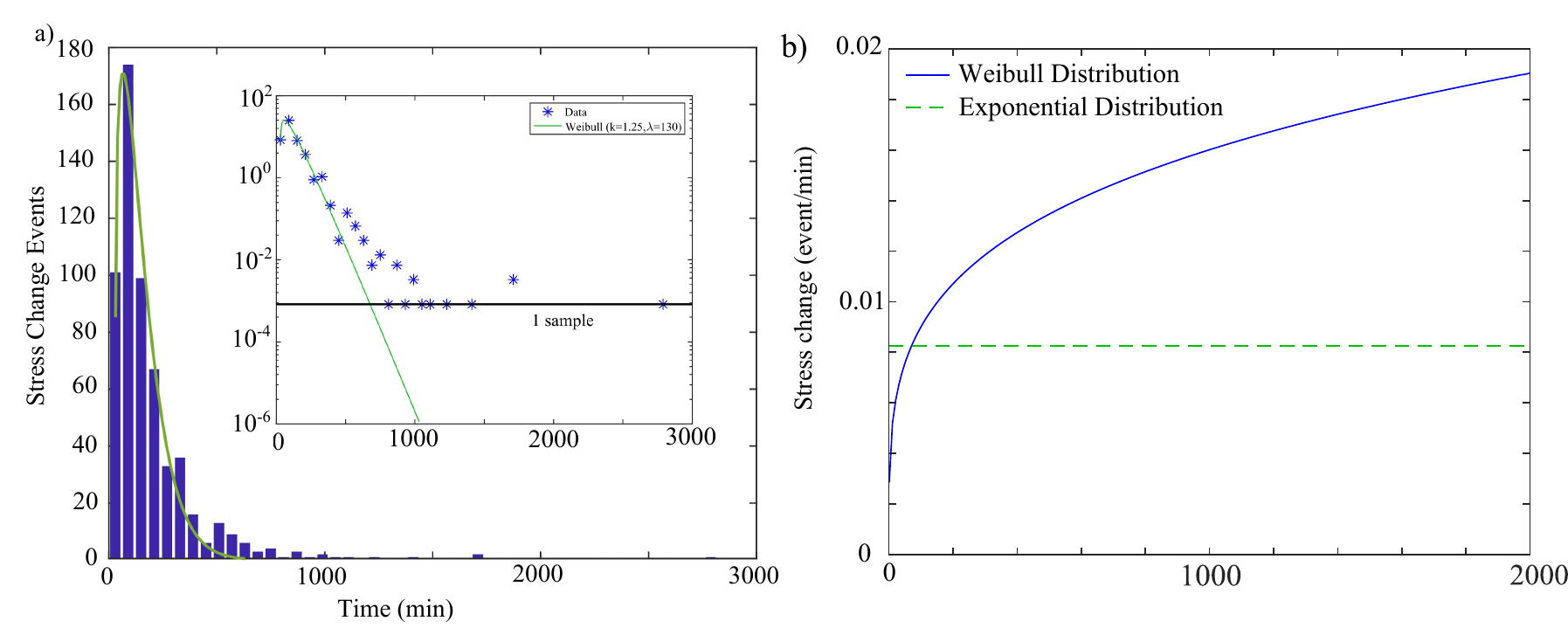}
    \caption{(a) The number of stress change events in a combination of 4 different experiments for a total of 34860 events.  The time between frames was 15 minutes for all experiments.  The fit is the PDF fitted with a Weibull distribution with a shape function $k=1.25$ and $\lambda=130$. There are several outliers to the distribution with much longer times between events (b) The chance of force chain movement increase over time 
    .}
    \label{fig:weibull}
\end{figure}
 
The roughly periodic perturbations to the system are set against a longer-term evolution in the force chains that roughly follows a Weibull distribution:

\begin{equation}
P(t)=\frac{k t^{k-1}}{\lambda^k}e^{\left(-\frac{t}{\lambda}\right)^k}
\end{equation}

The shape parameter $k$ of the Weibull distribution describes the rate at which contact stresses change.  A shape parameter $k>1$, as seen in the experiments in Figure \ref{fig:weibull}(a), indicate that the probability of failure increases with time (age weakening) and not regular perturbations. We are not directly measuring failure events, but rather stress rearrangements throughout the contact network and it should be noted that the perturbations are not growing with time in intensity or frequency to cause more rearrangements.  Redistribution of stress in the network could be considered failures (\textit{sensu stricto}), but determining the threshold of failure is difficult in these experiments as stress shifts (force-chain changes) can occur without particle movement.
 
\begin{figure}
    \centering
    \includegraphics[width=0.95\columnwidth]{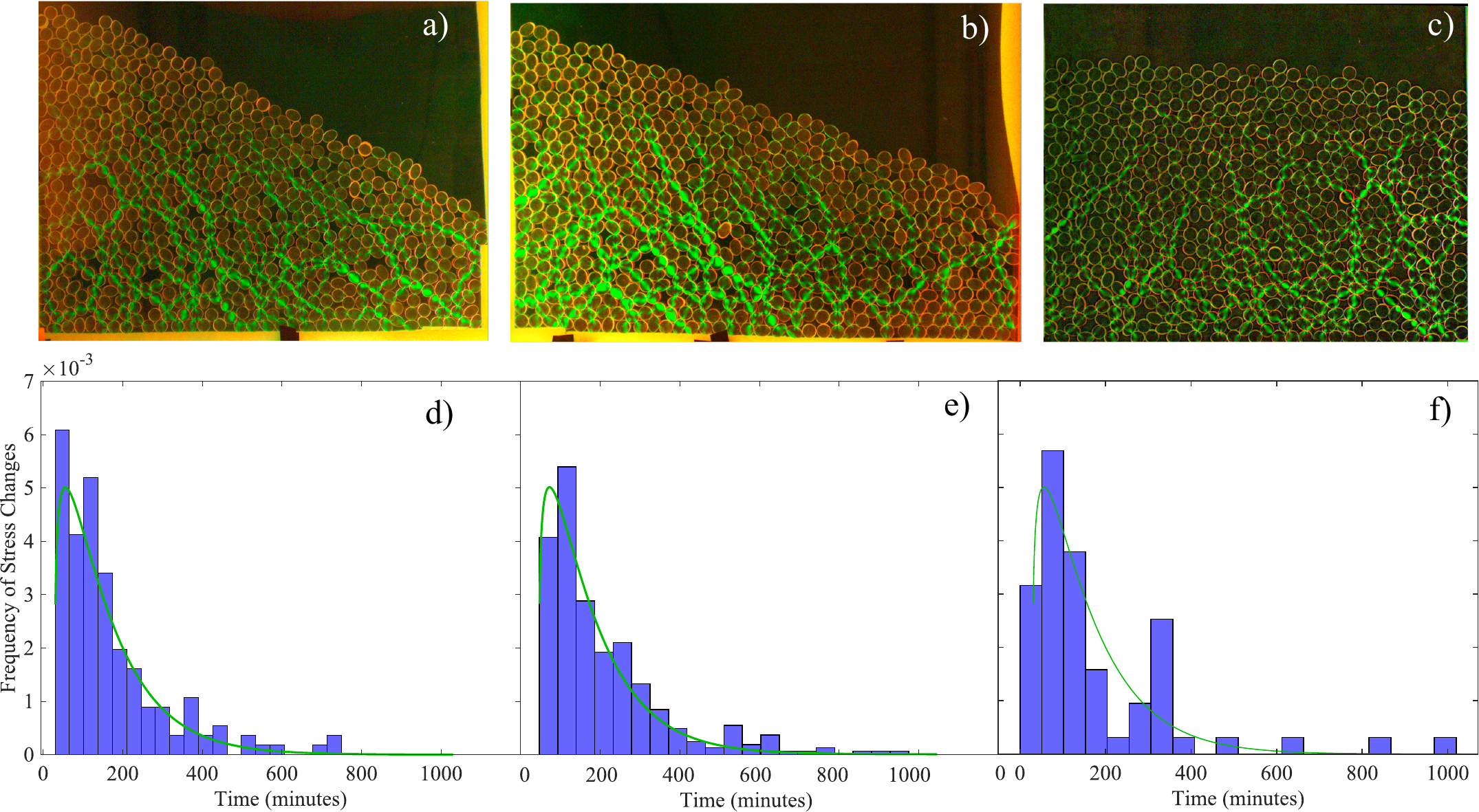}
    \caption{(a) Slope of 27.9$^o$, approximately the critical angle of repose (b) Slope of 19.2$^o$ (c) Slope of 3.5$^o$. (e-g) The corresponding force chain fluctuations with the conglomerate Weibull fit, $k=1.25$ and $\lambda=130$.}
    \label{fig:diffexp}
\end{figure}

The different surface slopes between experiments does not seem to greatly impact the failure distributions as shown in Figure \ref{fig:diffexp}, though slope angle does appear to affect the number of failure events. The time-weakening contact forces that determine the stress shifts appears to be roughly the same between different slopes and different stabilities.  The stability appears to change the amplitude of the probability of failure but the time dependence of the failure rate is similar across the different angles of repose.  This suggests a material property such as frictional weakening, as well as a relation of the stress imbalance/fragility, underlies the Weibull distribution. Even the minimal slope of grains shows stress relocation over long time periods  similar to that which is seen of an underwater granular bed \cite{allen2018}.

Following the initial settling events, changes in the pile occur at irregular intervals. In fact, even considering the contact weakening model of the Weibull distribution, there are still some events at long time scales that fall well outside of the distribution's expected failure. The longest time between stress rearrangements was calculated at around 46 hours (2760 minutes). Such long-range force-chain changes have a vanishing small probability with the fitted Weibull distribution, $P(x)=6.15\times 10^{-14}$. None of the experiments ran long enough to see a clear cessation of stress changes.  

\begin{figure}
    \centering
    \includegraphics[width=0.85\columnwidth]{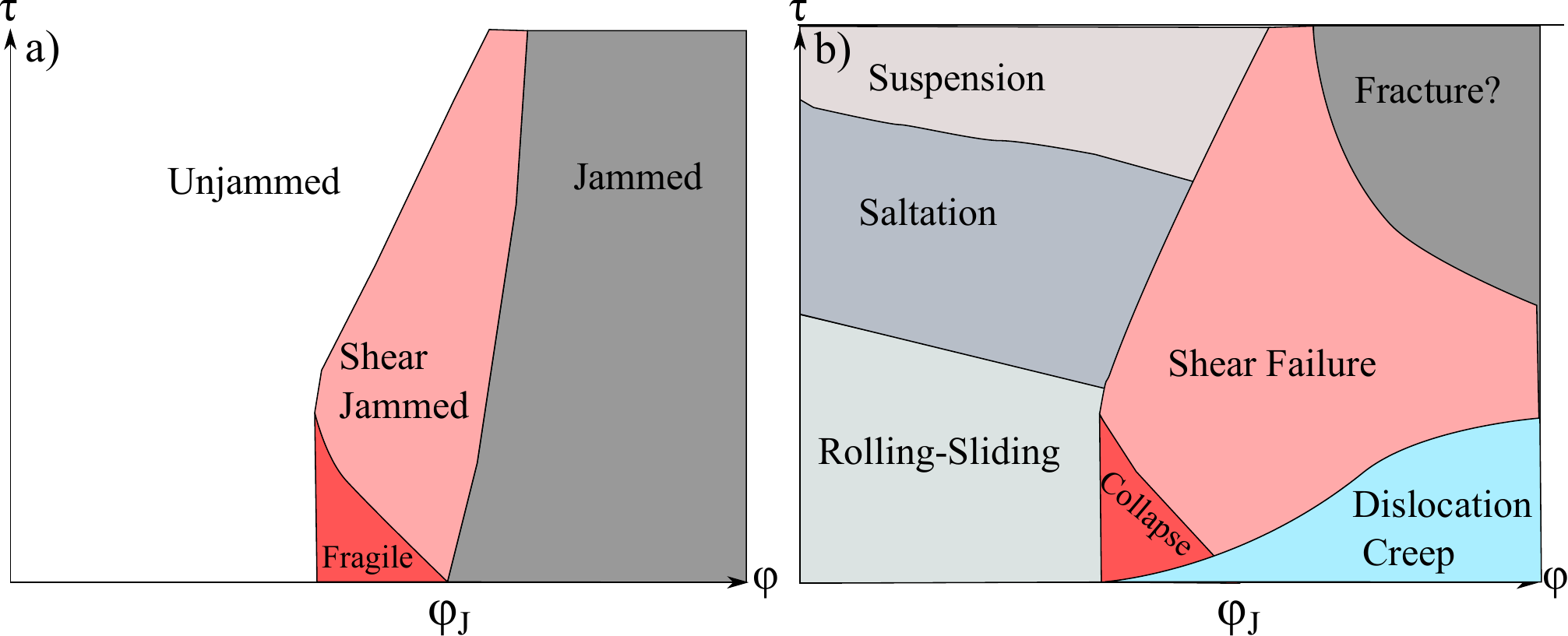}
    \caption{(a) Phase diagram of jamming showing the applied stress plotted against the packing fraction.  (b) Failure modes in the same phase space. At high applied stress, and low packing fraction the grains move and flow, at intermediate packing fraction they are partly jammed and resist movement in one direction but not others, and at high packing fractions the grains are solid and prone to more localized failure.}
    \label{fig:phasespace}
\end{figure}

Having described the time scales and energetics of force-chain changes and particle motions we turn now to a system-wide view of the granular piles. Firstly, any given slope of a granular pile is not at the granular material's global equilibrium because there is an uneven distribution of particles with different energies and stresses. The one exception would be a true crystalline structure obtained by careful packing \cite{panaitescu14} and a uniform distribution of stress for each layer \cite{hong1993}. In contrast with a crystallized system, natural granular piles exist in a fragile regime where long temporal and micro-rearrangements take place between the onset of system rigidity and full particle arrest\cite{lechenault2008,katsuragi2010}. Importantly, even in such a system not all energy changes can be measured by particle microslips\cite{johnson2013}, but as seen in these experiments there are energy fluctuations without particle movement.  For our experiments, the force-chain changes without corresponding particle movement is an expression of this fragile state. 

Another way to describe this fragile state of the slope, that is very relevant to this effort, is that the timescale of particle rearrangement grows towards infinity as the jammed state is approached. This rearrangement timescale is very different than the timescale needed to form a granular packing to its lowest energy state, or global equilibrium \cite{panaitescu14}. In all experiments we see evidence of the stresses localizing on a subset of contacts by the rising intensity in the force-chain network. Because there is a threshold in the amount of stress we can measure, we only are able to visualize stress above the weight of three particles. Therefore, if we are seeing a higher intensity in the force-chain color-space, and assume that the force is conserved across the experiment, the force chains are containing a growing percentage of the granular stress. The general trend suggests that the stress is localizing along the most stressed force chains until localized failure, as seen in Figure \ref{fig:apparatus}(b). Growing localized stress shows that the system is slowly weakening over time (i.e. becoming more fragile), as also indicated by the Weibull shape factor $k>1$.

We close with a brief consideration of how the experiment's fragile state, and the events that occur within it, can shed light on granular slope failure and creep in nature. Many  descriptions of landslides, earthquakes, and avalanches consider them as elastoplastic systems. Strain energy builds in the system via tectonic forces \cite{daniels2008}, pore-pressure variations as humidity or rainfall change \cite{iverson2005regulation}, and/or noise is imparted by weather events or human activities \cite{regmi2020}. As the system - the hillslope, the fault zone, the basal layer of the snowpack - reaches its strength threshold, failure occurs. Simple elastoplastic models can be adjusted with viscous terms \cite{hayman2014geologic} or velocity weakening frictional terms \cite{finnegan2024seasonal} depending on the environment. Such approaches go a long way to explaining anomalously weak and/or periodically slipping systems. In our experiments we wished to directly observe the forces within a granular pile in this context. 

In an effort to directly image forces in a system that has elsewhere been reported to sit in a granular-frictional, but glassy creep regime \cite{jerolmack2019viewing, deshpande2023sensitivity}, we observe that granular piles achieve fragile states that are never truly at rest. We observe migrations in stresses and displacements that are not simply explained by settling or external forcings of an elastoplastic system. The stress state of the pile does not simply fluctuate around a mean value, but rather migrates towards new configurations occasionally disturbing the particle arrangements causing small shifts. The resulting phase space, commonly expressed by packing density and mechanical terms of stress or strain rate, relates both fragile and jammed states, but also indicates how different deformation mechanisms ought to dominate as the system crosses any phase boundary away from fragility, as shown in Figure \ref{fig:phasespace}.
The unjammed space for example would cover a wide range of mechanisms ranging from rolling-sliding of grains within the pile or along the surface, to saltation (migration of grains over the top), or flows of suspension en masse, each marking an unjammed phenomena at respectively increasing density of the granular pack\cite{allen2018}. Alternatively, stronger jammed systems at higher applied stress will fail in a range from creep to shear failure to fracture; Figure \ref{fig:phasespace}b offers one possible range of behaviors for different jammed and unjammed states that build out from the fragile state in this experiment. 
The details of this phase space will take further research to implement the force-chain dynamics that are well imaged experimentally.

\section{Conclusion}
In an effort to experimentally probe failure and creep within a granular pile, we imaged force chain and particle rearrangements over month-long time periods. Imaged force chains, via the photoelastic properties, form an irregular network of contact stresses which is not static under gravity, but rather a localization of stresses to the strongest particle contacts until they fail. The force-chain evolution highlights discrete migration of the principle stress vectors over time. The closer the slope is to the angle of repose the more likely the stress is to shift and the particles to creep, but the time distribution of the stress changes is not substantially different between the different slopes, all of which are well below the critical angle of repose. Even after an initial settling time of about 5-6 hours contact forces shift and migrate. This force network migrates over time becoming steeper and more localized. Because the system is in a non-equilibrium fragile state, it is possible that the stresses are changing in response to mechanical changes on or within the slope above, or due to changes deeper within the pile. The stick-slip force network changes do not appear to be driven by slight mechanical noise, but there is some sensitivity to temperature changes of $\sim$ 2$^o$. The temperature changes likely affect the granular pile via microscopic thermal expansion and contraction within the particle configuration.   There is an increasing probability of a force-chain shifting with time as described by the Weibull distribution with shape factor $k > 1$. However, even after longer time scales, far along the tail of the Weibull distribution, there are long-time events that fall well outside the distribution.
However, the system is in a fragile state, with locally jammed regions that can cause force-chain changes and even displacements in other regions within the system. Such force-chain dynamics may govern natural systems that are in fragile states, but prone toward transitions into granular failure and creep over much longer time scales. 

\section{Acknowledgements}
We would like to acknowledge Karen Daniels for initial conversations about photoelasticity and the experimental methods.  We would also like to thank the Oklahoma Geologic Survey seismic team for the Raspberry Shake seismic monitor.

\section{Author Contributions}
B.A. performed the experiments, analysis and the original draft, N.H. contributed to conceptualizations and paper writing and editing.

\nocite{supplemental,daniels2017}

\raggedright
\bibliography{photoelastic}
\end{document}


\title{Force chain dynamics in a quasi-static granular pile}
\author{Benjamin Allen, Nicholas Hayman}
\affiliation{$1$ Oklahoma Geological Survey, University of Oklahoma, Norman OK 73071}

\date{\today}

\section{Calibration}
\begin{figure}
    \centering
    \includegraphics[width=0.85\columnwidth]{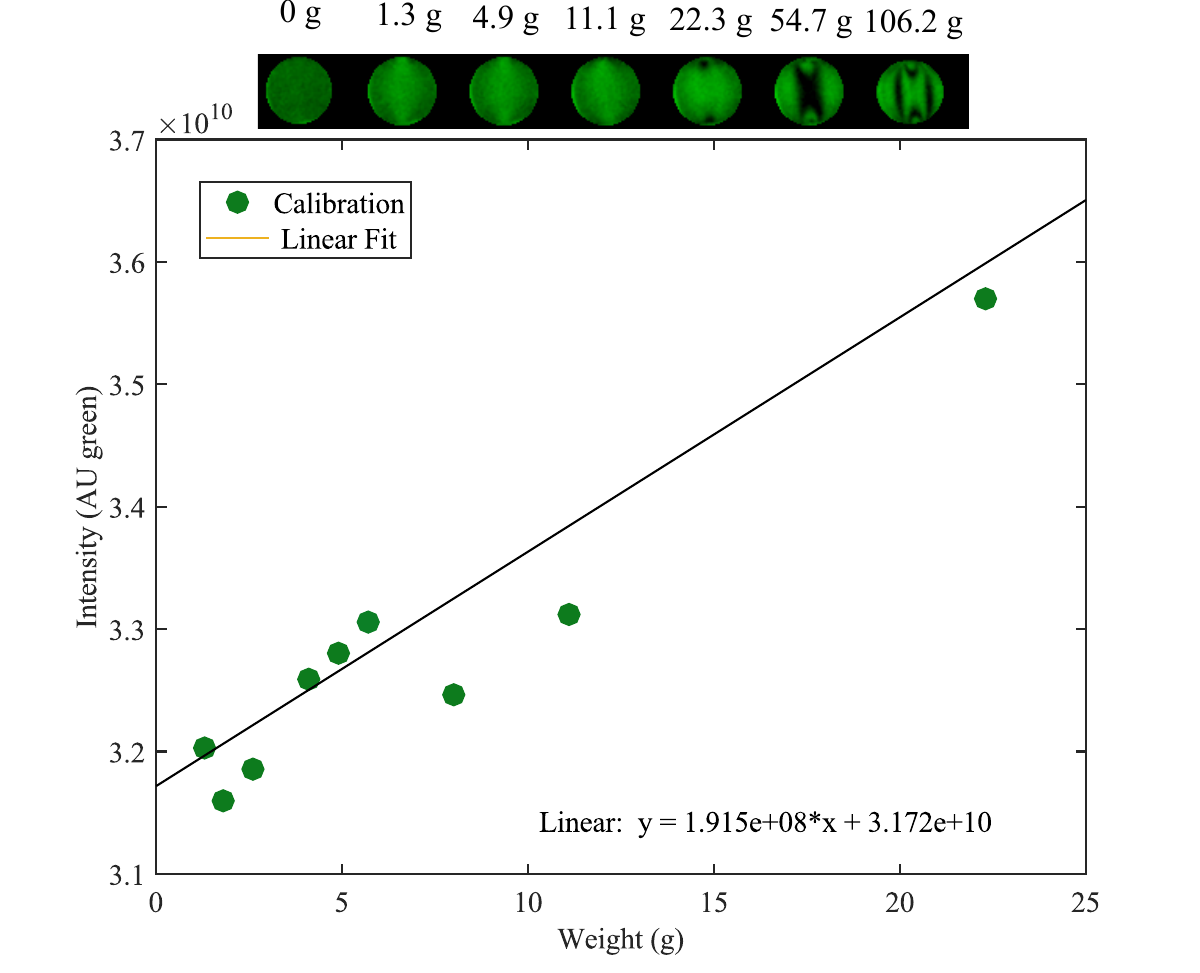}
    \caption{A calibration of the green light intensity squared $I$ of the photo-elastic effect in a single disc under different forces.  The relationship is essentially linear up to about $30\times m_{circle}g$.  Also to be noted is the number of fringes expanding around the contact points in the sample full-color images as the intensity-pressure relationship begins to deviate from linear.}
    \label{fig:G2calibration}
\end{figure}

Starting with a single particle and loading it up to about 35 times it's weight we see that the cross-polarized green light intensity initially increases linearly with the amount of loading shown in Figure \ref{fig:G2calibration}. The optimal polarization from the photo-elastic sheets is $\lambda =560$ nm.  Additionally, the threshold for stress detection is about the weight of two particles $I_o = I(2 w_{particle})$.  At approximately $F_n\sim 35 \times w_{circle}$, the linear relationship appears to level off and decrease as clear fringes become visible.  The stress calibration can be extended with counting the number of fringes across the particle horizontal to the line between the contacts, but within this experiment we stay within the linear regime with the height of the pile of particles being limited to about 25 particles.


Using a shadowgraph technique following a heat bath at $69^o$ C for half an hour we estimate the linear thermal expansion coefficient $\alpha \approx 0.00029 \pm .00005$ C$^{-1}$ in both radial and cylindrical direction.
The particles have a high thermal coefficient of expansion due to the photo-elastic particle being a soft rubber.

\subsection{Data Handling}
\begin{figure}
    \includegraphics[width=0.85\columnwidth]{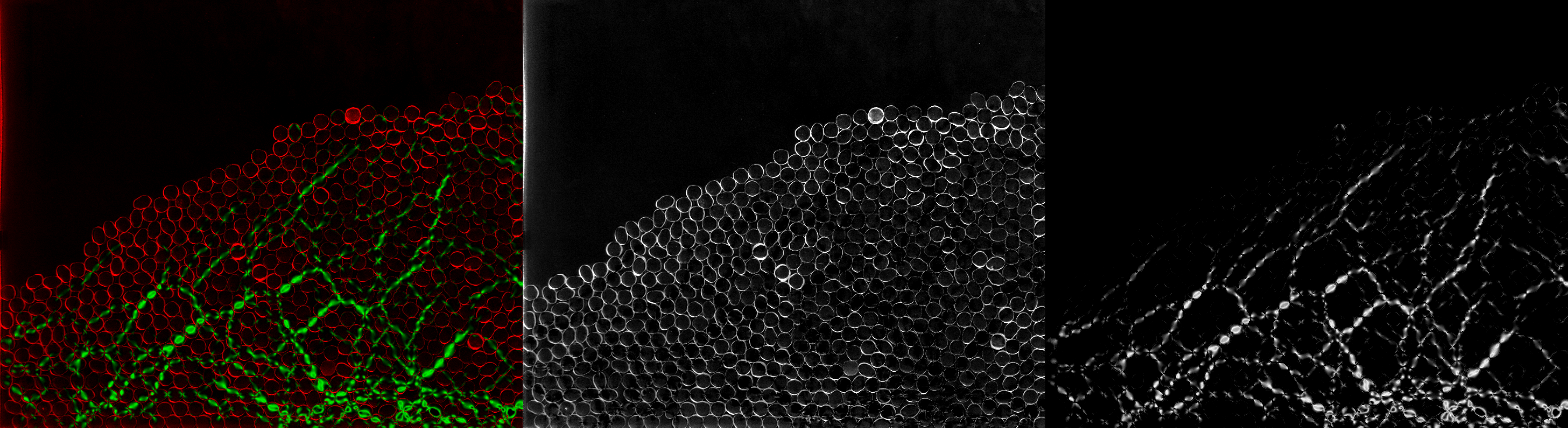}
    \caption{A representation of the deconvolution of the particle positions and force chains.}
    \label{fig:channels}
\end{figure}

Images were taken every 15 minutes and frontlit by red light and backlit by green.  The images were then split into red and green channels as shown in Figure \ref{fig:channels} to separate the force chains from the particle positions.  We attempted to filter out any extra scattered light and normalize the intensity across the image using the light from the apparatus frame.

The experiment was moved from one room to another in between experiments in June.

 During the data collection, cleaning staff turned on the light at approximately June 6 12:30 AM resulting in the removal of 2 data points where the light intensity spikes 2 orders of magnitude.

\begin{figure}
    \centering
    \includegraphics[width=0.85\linewidth]{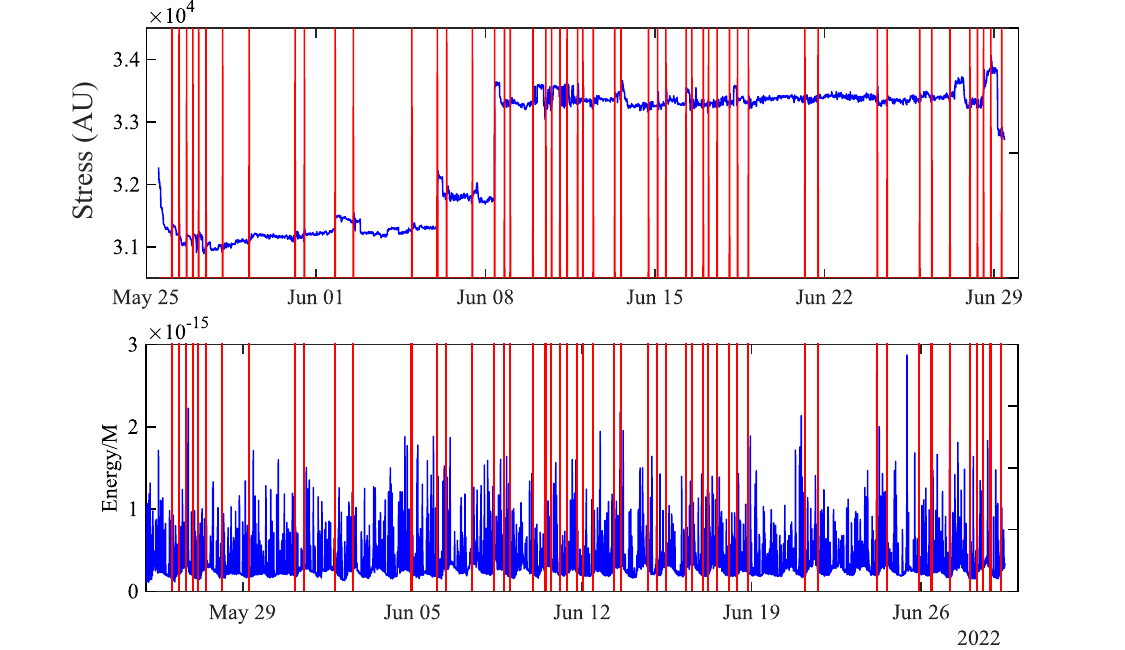}
    \includegraphics[width=0.85\linewidth]{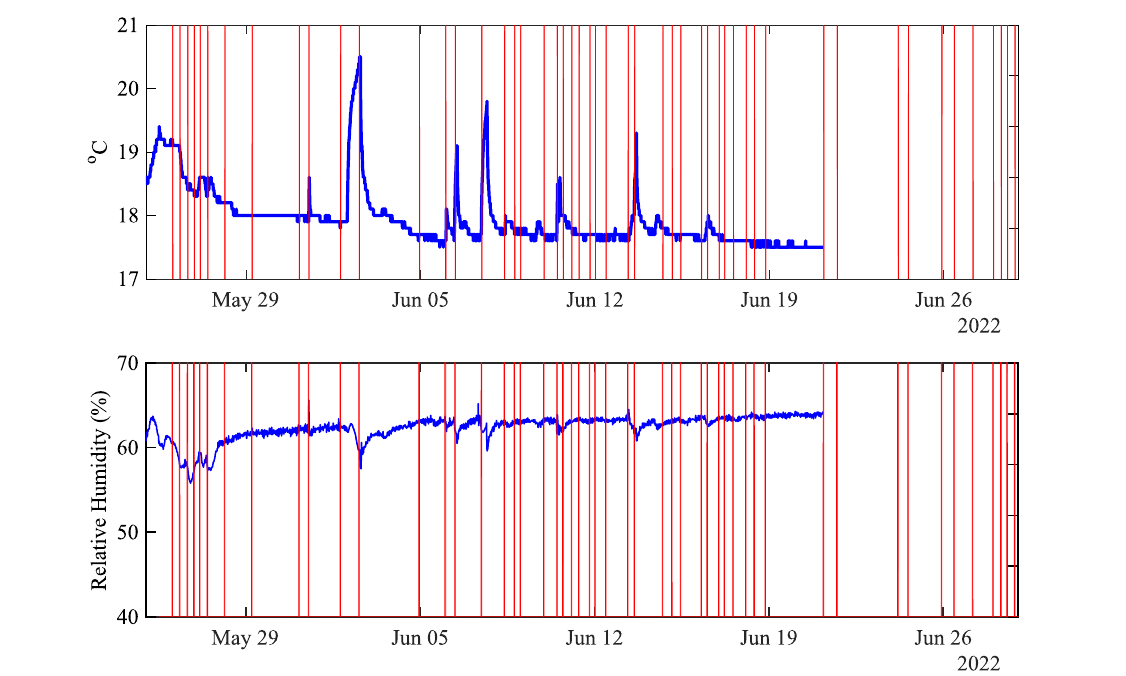}
    \caption{A breakdown of the force chain shifts and the outside stimuli during the longest experiment. Force chain shifts denoted by red lines.}
    \label{fig:shifts}
\end{figure}

 There was only one experiment where we have processed the seismic data and the temperature recordings with the results presented in the main manuscript and also shown in Figure \ref{fig:shifts} which denotes force chain shifts overlain on top of the outside stimuli.  The temperature-humidity sensor filled up with data 10 days before the experiment initially ended.
 
Due to the nature of resetting the experiment the lighting changes slightly between experiments.

\subsection{Boundary/Initial Conditions}

As with many granular experiments the preparation of the experiment can play a large role within determining the subsequent dynamics.  Granular materials are rarely close to their equilibrium conditions and appear to approach an average sampling that is called random packing when all the grains are constrained with balanced forces. This jammed state can not progress without changing the forces between particles or the environment.

This is a 2-D experiment using discs rather than natural grains due to the better control and measuring capabilities.  The phenomenon that we see within this experiment does not change when transferring to 3-D materials and structures, but the scaling might as there are more degrees of freedom.

Additionally, we do have boundaries to constrain the experiment on the sides. These boundaries are similar, but not completely equivalent, to having a free granular pile with a taper to the bottom surface and symmetry about the left side. The lack of equivalence can be seen just by looking at the force chain structure of the material created through disordered granular contacts comparing it to that of a solid wedge, but does not change any of the conclusions we draw from this experiment.

\subsection{Multiple Experiments}

\begin{figure}
    \centering
    \includegraphics[width=0.85\linewidth]{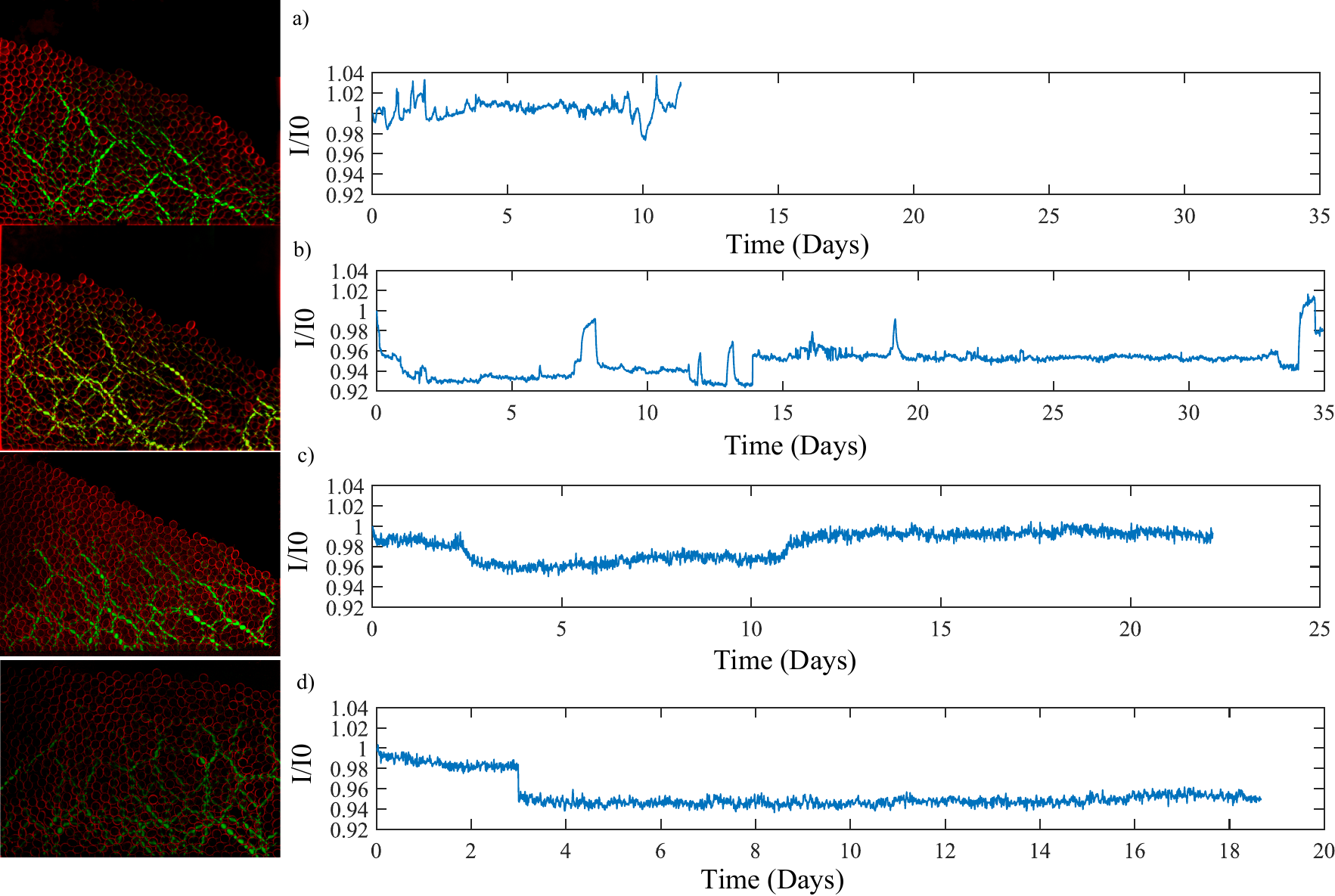}
    \caption{Experiments performed at different slopes (a) 24.5$^o$ (b) 27.5$^o$ (c) 26.5 $^o$ (d) 7$^o$ and the corresponding change in the force chain intensity $I$ normalized by the initial intensity $I0$.}
    \label{fig:expts}
\end{figure}

We discussed multiple experiments in the manuscript, the time series of force changes are laid out in Figure \ref{fig:expts}.  These experiments had different slopes, and unfortunately different lengths of time due to filling up disc space and maintenance.

\raggedright
\bibliography{photoelastic.bib}